\documentclass[conference]{IEEEtran}
\def\BibTeX{{\rm B\kern-.05em{\sc i\kern-.025em b}\kern-.08em
    T\kern-.1667em\lower.7ex\hbox{E}\kern-.125emX}}
\usepackage{cite}
\usepackage{caption}
\usepackage{subcaption}
\usepackage{graphicx}

\usepackage{amsmath}
\usepackage{amssymb}
\usepackage{amsthm}

\usepackage{stfloats}
\usepackage{bm}
\usepackage{algorithmic}
\usepackage{xcolor}
\usepackage{tikz}
\usetikzlibrary{shapes,arrows,positioning,plotmarks,shadows,calc,matrix,fit,patterns}
\usepackage{pgfplots}
\usepackage{todonotes}
\usepackage{datetime}
\usepackage{cancel}
\usepackage{hhline}
\usepackage{cases}
\usepackage{nicefrac}
\usepackage{etoolbox}
\usepackage{algorithm}
\usepackage{algorithmic}
\usepackage{cleveref}
\usetikzlibrary{plotmarks}
\usetikzlibrary{arrows.meta}
\usepackage{grffile}
\usepackage{dsfont}

\newcommand{\figref}[1]{Fig.~\ref{#1}}

\DeclareMathOperator{\re}{Re}
\DeclareMathOperator{\im}{Im}

\newcommand{\diag}{\text{diag}}

\newtheoremstyle{remarkmod}
  {\topsep}   
  {\topsep}   
  {\normalfont}  
  {0pt}       
  {\itshape} 
  {.}         
  {5pt plus 1pt minus 1pt} 
  {}          
\theoremstyle{remarkmod}

\newtheorem{corollary}{Corollary}
\newtheorem{theorem}{Theorem}
\newtheorem{lemma}{Lemma}
\newtheorem{remark}{\textit{Remark}}

\DeclareMathOperator*{\argmin}{argmin} 
\usepackage{graphicx}
\usepackage{mathtools, cuted}
\IEEEoverridecommandlockouts
\begin{document}
\title{{IRS-Assistance with Outdated CSI: Element subset selection for secrecy performance enhancement}\
\thanks{This work was funded by the Federal Ministry of Education and Research
	(BMBF) of the Federal Republic of Germany (Forderkennzeichen 16KISK095, 6G ANNA).}
}


\author{\IEEEauthorblockN{Chu Li, Aydin Sezgin}
	\IEEEauthorblockA{Ruhr-Universit\"at Bochum, Germany \\
		Email:  \{chu.li, aydin.sezgin\}@rub.de}
}

\maketitle

	\begin{abstract}
 	 In this work, we investigate the secrecy performance in an intelligent reflecting surface (IRS)-assisted downlink system. In particular, we consider a base station (BS)-side IRS and as such, the BS-IRS channel is assumed to be known perfectly. Of more importance, we consider the case, in which only outdated channel state information (CSI) of the IRS-user channel is available.  We study the impact of outdated CSI on the secrecy performance numerically and analytically. Furthermore, we propose an element subset selection (ESS) method in order to improve the secrecy performance. A key observation is that minimal secrecy outage probability (SOP) can be achieved using a subset of the IRS, and the optimal number of selected reflecting elements can be effectively found by closed-form expressions. 
\end{abstract}


\section{Introduction}
%
Due to the broadcast nature of the wireless medium, transmission over a wireless network is prone to eavesdropping on information intended to be exchanged between legitimate terminals. {To securely transmit the data, physical security layer (PLS) techniques can be applied \cite{bloch2008wireless,7849064}. }In recent years, intelligent reflecting surfaces (IRS) have been applied in the field of PLS, which shows great potential to improve the security of wireless communication. IRS is a surface consisting of a large number of reflecting elements whose phase can be adaptively controlled by a microcontroller. The basic idea of an IRS-assisted system is to configure the IRS to reflect the signal in the direction of the desired legitimate receiver. Compared to conventional systems, IRS-assisted systems can achieve higher reliability and security at a lower cost \cite{bjornson2022reconfigurable,staat2022irshield, 9759225}. 

The secrecy performance of IRS-assisted systems has been studied in many recent works. In \cite{illi2022secrecy,trigui2021secrecy}, the secrecy performance is analyzed in terms of secrecy outage probability (SOP) and secrecy rate in single-input single-output (SISO) systems considering the quantized phase error at IRS. The authors in \cite{niu2021weighted} jointly designed the secure beamforming and artificial noise (AN) to maximize the secrecy rate in IRS-assisted multiple-input single-output (MISO) systems using the alternating direction method of multiplier (ADMM). This work is later extended to multiple-input multiple-output (MIMO) systems in \cite{chu2020secrecy}, where majorization-minimization (MM) is used. The aforementioned works all assume perfect CSI, which is usually not the case in practice. In \cite{yu2020robust}, the authors proposed a robust design of the beamformer at the base station (BS) and the phase shifters at IRS to maximize the system sum rate considering CSI with estimation errors. Besides the estimation error, outdated CSI is also a major contributor to CSI imperfections \cite{ferdinand2013effects}.  In practice, the channel often changes with time. In addition, the channel estimation process and configuration of the beamformer at the BS and the phase shifters at the IRS may take time, especially if the number of reflecting elements is large.  Consequently, the CSI observed by the BS may be outdated for subsequent data transmission. Configuring the beamformer at the BS and IRS phase shifters using outdated CSI will result in a loss of signal-to-noise ratio (SNR) and may degrade the secrecy performance.

In this work, we study the secrecy performance in IRS-assisted downlink systems considering outdated CSI of IRS-user channel. Since the phase shift at IRS cannot be configured precisely in practice, we further assume that each reflecting element suffers from an independent and uniformly distributed random phase error. We present the statistical characterization of SNR at Bob and Eve and derive the closed-form expressions of the SOP taking into account the outdated CSI. In addition, we propose a novel element subset selection (ESS) method that has low complexity and can be used to improve secrecy performance. More specifically, $K$ of the total $N$ reflecting elements are selected and turned on during transmission, while the other elements are turned off. The correctness of the closed-form expressions is verified by Monte-Carlo simulations. Additionally, we observe that the minimum SOP is achieved by using a subset of the reflecting elements.

\section{System Model}
\label{sec:chmod}


In this work, we consider a BS with $M$ antennas communicating with a legitimate user Bob in the presence of an eavesdropper Eve. Both Bob and Eve are equipped with  single antennas. 
  The transmission is facilitated by IRS, which consists of $N = N_H N_V$ reflecting elements. The size of each element is $d_H \times d_V$, where $d_H$ and $d_V$ are the horizontal width and vertical height, respectively. Herein, we consider the IRS at the BS side, i.e., both BS and IRS are on the top of some high-rise buildings and close to each other \cite{you2022deploy}. Meanwhile, the IRS can be adaptively configured by the BS. Considering BS-side IRS, the channel between BS and IRS is modeled as
\begin{align}
  	\label{eq:BS_IRS_ch}
    \mathbf{H} = \sqrt{\beta_{H}}\mathbf{a}({\varphi_1, \theta_1})\mathbf{b}^H({\varphi_2, \theta_2}), 
\end{align} 
where $\beta_{H}$ is the distance-dependent path loss factor, $\mathbf{a}(\varphi_1, \theta_1) \in \mathbb{C}^{M} $ and $\mathbf{b}(\varphi_2, \theta_2) \in \mathbb{C}^{N}$ are the steering vectors at BS and IRS, respectively. In this context, $\varphi_1$ and $\theta_1$ represent the azimuth and elevation angle of departure (AoD) at BS, while $\varphi_2$ and $\theta_2$ denote the azimuth and elevation angle of arrival (AoA) at IRS. Furthermore, the $m$-th element of $\mathbf{a}$ is given by   
\begin{align}
	\label{eq:sv_BS}
	\mathbf{a}_m = e^{j2\pi \frac{d_{BS}}{\lambda}(m-1)\sin\varphi_1 \sin\theta_1},\end{align}
where $d_{BS}$ and $\lambda$ are the inter-antenna separation at the BS and the carrier wavelength, respectively. Also, the $n$-th element of  $\mathbf{b}$ is
\begin{align}
	\label{eq:sv_IRS}
	\mathbf{b}_n = e^{j \frac{2\pi}{\lambda}(k(n)d_H\cos\theta_2 \sin\varphi_2+l(n)d_V\sin\theta_2) }, 
\end{align}
where $k(n)=\mod \left(n-1, N_{\mathrm{H}}\right)$ and $l(n)=\left\lfloor(n-1) / N_{\mathrm{H}}\right\rfloor$ are the horizontal and vertical indices of $n$-th element, respectively \cite{bjornson2020rayleigh}. In this work, it is assumed that $d_{BS}=d_H = d_V = 0.5 \lambda$, such that there is no correlation between the individual transmit antennas and the reflecting elements\footnote{Although we set $d_H = d_V = 0.5 \lambda$, there is still a weak correlation between the reflecting elements. However, the effect of such a weak correlation is negligible, which can be easily demonstrated by simulations. Therefore, in this work, it is assumed that there is no correlation between the individual reflecting elements}.
In practice, the distance-dependent path loss factor $\beta_{H}$ in \eqref{eq:BS_IRS_ch} remains unchanged for a long time. Also, $\mathbf{a}(\theta_{\varphi_1, \theta_1})$ and $\mathbf{b}(\theta_{\varphi_2, \theta_2})$ can be calculated accurately if the location and construction of the IRS are known. For this reason, we assume that perfect information of $\mathbf{H}$ is available at {BS, Bob and Eve}.
Furthermore, we use a $N \times N$ diagonal matrix to represent the IRS
\begin{align}
	\label{eq:IRS_Ps_1}
	\tilde{\mathbf{\Phi}} = \diag{(e^{j\tilde{\phi}_{1}}, e^{j\tilde{\phi}_{2}},  \cdots,  e^{j\tilde{\phi}_{N}} )},
\end{align}
where $\tilde{\phi}_{n} = {\phi}_{n} +\Delta{\phi}_{n}$ is the phase shift at the $n$-th reflecting element, $ {\phi}_{n}$ and $\Delta{\phi}_{n}$ are the precise phase shift and random phase error at each element, respectively. More specifically, here we consider the discrete phase shift in IRS such that $\Delta{\phi_{n}}$ is uniformly distributed in $[-\frac{\pi}{L}, \frac{\pi}{L}]$, where $L$ is the number of quantization levels \cite{badiu2019communication}. Moreover, the characteristic function of $\Delta{\phi_{n}}$ can be computed by
\begin{align}
	\label{eq:character_f}	
	\mu_p = \mathbb{E}[e^{jp\Delta{\phi_{n}}}] = si\left( -p\frac{\pi}{L}\right) , \qquad p \in \mathbb{Z},
\end{align} 
where $si(x) = \frac{\sin(x)}{x}$.
The IRS-Bob channel and IRS-Eve channel are modeled as 
\begin{align}
	\label{eq:IRS_UE_ch}
	\mathbf{h}_B  \sim \mathcal{CN}(0, \beta_{B} \mathbf{I}_N),  \qquad
	\mathbf{h}_E  \sim \mathcal{CN}(0, \beta_{E} \mathbf{I}_N), 
\end{align} 
where $\beta_{B}$ and $\beta_{E}$ are the path-loss factors. Furthermore, the direct link is assumed to be blocked by obstacles. Hence, the received signal at Bob and Eve are, respectively, given by
\begin{align}
	\label{eq:y_B}	
	y_B& = \sqrt{P} {( \mathbf{H}	\tilde{\mathbf{\Phi}} {	\mathbf{h}}_B) ^H \boldsymbol{w} } x + n_B, \\
	\label{eq:y_E}
	y_E& =\sqrt{P} {( \mathbf{H}	\tilde{\mathbf{\Phi}} {	\mathbf{h}}_E) ^H \boldsymbol{w}} x + n_E,
\end{align}
where $P$ is the transmit power, $\boldsymbol{w}$ denotes the beamformer at BS, $x$ is the transmit data symbol with $\mathbb{E}[\left|x \right| ^2]=1$, $n_B \sim \mathcal{CN}(0, \sigma_{B}^2)$ and $n_E \sim \mathcal{CN}(0, \sigma_{E}^2)$ are the receiver noise at Bob and Eve, respectively.

In this work, we consider a practical passive eavesdropper where the instantaneous CSI of $\mathbf{h}_E$ is not available at the BS. { Note that, unlike the BS-IRS channel, we assume that the instantaneous CSI of $\mathbf{h}_B$ observed by the BS is outdated.  This is because the channels between the IRS and the users usually changes over time, caused by the movements of the users and other objects around them. }
  Let $\hat{\mathbf{h}}_B$ denote the outdated version of $\mathbf{h}_B$, the relation between $\hat{\mathbf{h}}_B$  and ${\mathbf{h}}_B$ can be expressed as 
\begin{align}
	\label{eq:dynamic_h_B}
	\mathbf{h}_B = \rho  \hat{\mathbf{h}}_B + \mathbf{e}_B, 
\end{align}
where $\rho = J_0(2\pi f_d T_d)$ is the correlation coefficient, and $J_0$ is the zero-order Bessel function, $f_d$ and $T_d$ are the Doppler frequency and the time distance between $\mathbf{h}_B$ and $\hat{\mathbf{h}}_B$, respectively. Moreover, $\mathbf{e}_B$ is the complex Gaussian distributed error with covariance matrix $(1-\rho^2)\beta_{g} \mathbf{I}_N$. The beamformer $\boldsymbol{w}$ and IRS will be designed {based on} $\mathbf{H}$ and $\hat{\mathbf{h}}_B$. More specifically, we apply maximum ration transmission (MRT) to design the beamformer at BS expressed as
\begin{align}
	\label{eq:beam_BS}
	\boldsymbol{w} = \frac{\mathbf{H}	{\mathbf{\Phi}}	 \hat{\mathbf{h}}_B}{ \big\|\mathbf{H}	{\mathbf{\Phi}}	\hat{\mathbf{h}}_B \big\| }.	
\end{align}
Now, similar to \eqref{eq:dynamic_h_B}, we use $\hat{\mathbf{h}}_E$ to represent the outdated version of $\mathbf{h}_E$ modeled as
\begin{align}
	\label{eq:dynamic_h_E}
	\mathbf{h}_E = \rho  \hat{\mathbf{h}}_E + \mathbf{e}_E, 
\end{align}
where $ \mathbf{e}_E \sim \mathcal{CN}(0, (1-\rho^2)\beta_{E} \mathbf{I}_N)$. {Based on the system model introduced in this section, we analyze the secrecy performance in the following sections. }

\section{Secrecy Analysis}
\label{sec:Secrecy Analysis}
Let $\gamma_{B}$ and $\gamma_{E}$ denote the instantaneous SNR at Bob and Eve, respectively. According to \cite{bloch2008wireless}, the instantaneous secrecy capacity is defined as
\begin{align}
	\label{eq:sec_cap}	
	C_s(\gamma_{B},\gamma_{E}) = \left[ \log_2 (1+\gamma_{B})-\log_2 (1+\gamma_{E})\right] ^+,
\end{align}
where $[x]^+ = \max(0,x)$. { Unlike most state-of-the-art works, where Bob and Eve have their own perfect CSI, we investigate the secrecy performance in different scenarios, 1) Bob (and Eve) know their individual outdated CSI $\hat{\mathbf{h}}_{B}$ (and $\hat{\mathbf{h}}_{E}$), 2)  Bob (and Eve) know their individual perfect CSI ${\mathbf{h}_{B}}$ (and ${\mathbf{h}_{E}}$), } 3) Bob knows his outdated CSI $\hat{\mathbf{h}}_{B}$ while Eve knows her perfect CSI ${\mathbf{h}_{E}}$. In the following, we characterize the received SNR at Bob and Eve in \ref{sec:Received_SNR_at_Bob_Eve} and \ref{sec:SNR_Eve}. The secrecy outage probability is investigated in \ref{sec:Secrecy_performance}.
\subsection{SNR at Bob}
\label{sec:Received_SNR_at_Bob_Eve}
In this subsection, we analyze the SNR at Bob assuming outdated and perfect CSI, respectively. 
If Bob knows his outdated CSI, i.e., $\hat{\mathbf{h}}_B$, and coherent detection is performed, the received signal at Bob in \eqref{eq:y_B} can be recast as 
\begin{align}
	\label{eq:y_B1}	
	y_B& = \sqrt{P} { (\mathbf{H}{\mathbf{\Phi}} {	\mathbf{h}}_B)^H \boldsymbol{w} } x + n_B \nonumber \\ 
	&=\sqrt{P} {\big(\mathbf{H}	\tilde{\mathbf{\Phi}} ( \rho {	\hat{\mathbf{h}}_B} + {\mathbf{e}}_B )\big)^H \boldsymbol{w} } x + n_B \nonumber \\ 
	& = \sqrt{P} \rho { ( \mathbf{H}	\tilde{\mathbf{\Phi}}  {	\hat{\mathbf{h}}_B})^H \boldsymbol{w}  }x +  \underbrace{\sqrt{P} { (\mathbf{H}	\tilde{\mathbf{\Phi}}{\mathbf{e}}_B)^H  \boldsymbol{w} } x + n_B }_{\hat{n}_B}
	.
\end{align} 
Here, we treat the term containing the random error ${	\mathbf{e}}_B$ as the additional noise, and we use $\hat{n}_B$ to denote the overall effective noise.
The variance of $\hat{n}_B$ is calculated as 
\begin{align}
	\label{eq:n_B_tilde_var}
	\hat{	\sigma}_{B}^2 &= \mathbb{E}[  \hat{	{n}}_B  \hat{{n}}_B^H] = \sigma_B^2+ P \mathbb{E} \left[ \bigg|  (\mathbf{H}	\tilde{\mathbf{\Phi}}  {	\mathbf{e}}_t)^H \frac{ \mathbf{H}	{\mathbf{\Phi}}	\hat {\mathbf{h}}_B}{\|\mathbf{H}	{\mathbf{\Phi}}	\hat {\mathbf{h}}_B \| }\bigg| ^2 \right]  \nonumber \\ 
	& = \sigma_B^2+ PMN\beta_{H}(1-\rho^2)\beta_{B},
\end{align}
where the last equality is obtained by substituting \eqref{eq:BS_IRS_ch} and \eqref{eq:beam_BS}.
Thus, the instantaneous SNR at Bob under the assumption of outdated CSI is given by
\begin{align}
	\label{eq:gamma_B_hat}		
		\hat{\gamma}_B& = \frac{P}{\mathbb{E}[ \hat{	\mathbf{n}}_B \hat{	\mathbf{n}}_B^H]} \big|\rho ( \mathbf{H}	\tilde{\mathbf{\Phi}}  {	\hat{\mathbf{h}}_B} )^H \boldsymbol{w} \big|^2 \nonumber \\ 
	& = \frac{P}{	\hat{	\sigma}_{B}^2 } M \beta_{H} \rho^2 \left|  \textstyle \sum_{n=1}^{N} \hat{\mathbf{h}}_{B,n}^H   e^{-j \tilde{\phi}_n} \mathbf{b}_n  \right|^2,
\end{align}
where $\mathbf{b}_n$ and $ \hat{\mathbf{h}}_{B,n}$ are the $n$-th element of $\mathbf{b}$ and $ \hat{\mathbf{h}}_{B}$, respectively. If Bob has his perfect channel information, i.e., ${\mathbf{h}}_{B}$, the instantaneous SNR at Bob is given by
\begin{align}
	\label{eq:gamma_B_tilde}		
	\tilde{\gamma}_B
	& = \frac{P}{\sigma_B^2} M \beta_{H}  \left| \textstyle \sum_{n=1}^{N} {\mathbf{h}}_{B,n}^H   e^{-j \tilde{\phi}_n} \mathbf{b}_n \right|^2.
\end{align}
Next, we design the IRS to maximize the secrecy capacity in \eqref{eq:sec_cap}. {To this end, }the optimal IRS should be designed to maximize the SNR at Bob and minimize the SNR at Eve. However, since we assume that BS is unaware of Eve's instantaneous CSI, we find the optimal IRS by maximizing the SNR at Bob. 
In this context, the optimal IRS is obtained as \vspace{-1em}
\begin{align}
	\label{eq:opt_IRS}
	\phi_n^{opt} = \angle  \left( \hat{\mathbf{h}}_{B,n}^H  \mathbf{b}_n\right) .
\end{align}
It is noticed that the optimal IRS maximizing $\hat{\gamma}_{B}$ in \eqref{eq:gamma_B_hat} and $\tilde{\gamma}_{B}$ in \eqref{eq:gamma_B_tilde} are the same since only $ \mathbf{H}$ and $\hat{\mathbf{h}}_{B}$ are known to the BS and the IRS is being controlled by the BS.  By plugging \eqref{eq:opt_IRS} in \eqref{eq:gamma_B_hat} and \eqref{eq:gamma_B_tilde}, we observe the distributions of	$\hat{\gamma}_{B} $ and $\tilde{	\gamma}_{B}$ in the following lemma. 
\begin{lemma}
	$\hat{\gamma}_{B} $ and $\tilde{	\gamma}_{B}$ follow the Gamma distributions, \vspace{-0.3em}
	\begin{align}
		\label{eq:rec_SNR_dis}	
		\hat{\gamma}_{B} \sim \mathrm{Gamma} \left(  \hat{\kappa}_B, \hat{\omega}_B \right),  \\
		\tilde{	\gamma}_{B} \sim \mathrm{Gamma} \left(  \tilde{\kappa}_B, \tilde{\omega}_B \right),
	\end{align}
{where the scale parameters $\hat{\kappa}_B$, $\tilde{\kappa}_B$, and the shape parameters $\hat{\omega}_B$,  $\tilde{\omega}_B$ are introduced in Appendix \ref{sec:proof_lemma_1}. }
 The corresponding average SNRs at Bob are given by  \vspace{-0.3em}
\begin{align}
	\label{eq:average_gamma_B_out2}
	\mathbb{E}[\hat{\gamma}_B] &=  \frac{P}{	\hat{	\sigma}_{B}^2 } M \beta_{H}(\rho^2N\beta_{B}+\rho^2N(N-1)\frac{\pi}{4}\beta_{B}\mu_{1}^2), \\
	\label{eq:average_gamma_B_perfect}
	\mathbb{E}[\tilde{\gamma}_B] &= \frac{P}{{	\sigma}_{B}^2 } M \beta_{H}(N\beta_{B}+\rho^2N(N-1)\frac{\pi}{4}\beta_{B}\mu_{1}^2).
\end{align} 
\end{lemma}
%
\begin{proof}
	See Appendix \ref{sec:proof_lemma_1}.
\end{proof}	

\subsection{SNR at Eve}
\label{sec:SNR_Eve}
Now, we investigate the SNR at Eve. Similar to \eqref{eq:gamma_B_hat} and \eqref{eq:gamma_B_tilde}, the instantaneous SNR at Eve assuming outdated and perfect CSI are, respectively, given by
 \begin{align}
 	\label{eq:gamma_E_hat}		
 		\hat{\gamma}_E& = \frac{P}{\hat{\sigma}_{E}^2} M \beta_{H} \rho^2 \big|  \textstyle \sum_{n=1}^{N}  \hat{\mathbf{h}}_{E,n}^H e^{-j \tilde{\phi}_n} \mathbf{b}_n  \big|^2, \\
	\label{eq:gamma_E_tilde}		
	\tilde{\gamma}_E& = \frac{P}{\sigma_{E}^2} M \beta_{H} \big| \textstyle \sum_{n=1}^{N}  {\mathbf{h}}_{E,n}^H e^{-j \tilde{\phi}_n} \mathbf{b}_n \big|^2,
\end{align}
where 	\begin{align}
	\hat{\sigma}_{E}^2 =\sigma_E^2+ PMN\beta_{H}(1-\rho^2)\beta_{E}. 
\end{align}
{Next, we obtain the distribution of Eve's SNR as follows. }
\begin{lemma}
	$\hat{\gamma}_{E} $ and $\tilde{\gamma}_{E} $ follow the exponential distributions,	 
	\begin{align}
		\hat{\gamma}_{E} \sim \operatorname{Exp}(\hat{\lambda}_E),  \qquad 
		\tilde{\gamma}_{E} \sim \operatorname{Exp}(\tilde{\lambda}_E),
	\end{align}
where  $\hat{\lambda}_E = \mathbb{E}[\hat{\gamma}_E] = \frac{P M \beta_{H}  \rho^2N {\beta}_{E}}{\sigma_E^2+ PMN\beta_{H}(1-\rho^2)\beta_{E}}$ and $\tilde{\lambda}_E = \mathbb{E}[\tilde{\gamma}_E] = \frac{P}{\sigma_{{E}}^2} M \beta_{H} N {\beta}_{E}$.
\end{lemma}
\begin{proof}
The IRS is designed according to IRS-Bob channel, thereby $\phi_{n}^{opt}$ can be considered as a randomly distributed variable by Eve. Thus, the SNRs at Eve can be approximated to exponential distributed variables \cite{trigui2021secrecy}.
\end{proof}

%

\subsection{Secrecy outage probability (SOP)}
\label{sec:Secrecy_performance}
In this subsection, we analyze the secrecy performance in terms of the SOP, which is a crucial metric for measuring the security of a wireless channel. The SOP is defined as the probability that the secrecy capacity falls below a certain secrecy rate $R_{s}$,  
\begin{align}
	\label{eq:SOP}	
	P_{so} &= \operatorname{Pr} (	C_s\leqslant R_{s})
		=\operatorname{Pr}\left[\log _{2}\left(\frac{1+\gamma_{{B}}}{1+\gamma_{{E}}}\right) \leqslant R_{{s}}\right] \nonumber \\
		&=\operatorname{Pr}\left[\gamma_{{B}} \leqslant 2^{R_{{s}}}\left(1+\gamma_{{E}}\right)-1\right] \nonumber \\ 	
		&= \int_{0}^{\infty} F_{\gamma_B} (2^{R_{{s}}}\left(1+\gamma_{{E}}\right)-1) f_{\gamma_{{E}}}(\gamma_E) d \gamma_{{E}}, 
\end{align}
where $\gamma_{B} \in [\hat{\gamma}_B,\tilde{\gamma}_B ]$, and $\gamma_{E} \in [\hat{\gamma}_E,\tilde{\gamma}_E ]$. $F_{\gamma_B}(\cdot)$ and $f_{\gamma_E}(\cdot)$ are the corresponding probability density function (pdf) and cumulative density function (cdf) of the SNR at Bob and Eve.
\begin{theorem}
	The exact SOP is given by  
	\begin{align}	
		\label{eq:SOP_opt}	
		P_{so} =& \frac{1}{\Gamma\left(\kappa_B\right)} \textstyle \sum_{p=0}^{\infty} \frac{\left(-\frac{1}{{\omega}_{B}}(2^{R_{\mathrm{s}}}-1)\right)^{p}}{p!} \nonumber \\
		& \times \mathrm{G}_{3,3}^{2,2}\left[ \frac{ {\omega}_{B}}{ {\lambda}_{E} 2^{R_{\mathrm{s}}} }\mid \begin{array}{c}
			1,1+p-\kappa_B, 1+p \\
			1,p, 1+p
		\end{array}\right],
	\end{align}
	where $\kappa_B \in [\hat{\kappa}_B, \tilde{\kappa}_B ]$, ${\omega}_{B}  \in [\hat{\omega}_{B}, \tilde{\omega}_{B}]$, ${\lambda}_{E}  \in [\hat{\lambda}_{E}, \tilde{\lambda}_{E}]$ and $\mathrm{G}_{\cdot,\cdot}^{\cdot,\cdot}$ denotes the Meijer’s G-function \cite{gradshteyn2014table}.  {Note that even though \eqref{eq:SOP_opt} contains infinite series, it converges quickly and can be approximated by a few terms.}
%
\end{theorem} 
\begin{proof}
The proof idea is analogous to \cite[Corollary 9]{badarneh2020achievable}, hence omitted due to space constraints.
\end{proof}
{To obtain further insights on the SOP, the following corollary is provided.}
\begin{corollary}
The lower bound of the SOP is obtained as  
\begin{align}
	\label{eq:SOP_opt_low}	
	P_{so}	 
	\geqslant \left( \frac{ {\lambda}_E}{{2^{-R_{\mathrm{s}}}{\omega}_B} +  {\lambda}_E}\right) ^{\kappa_B},
\end{align}
where $\kappa_B \in [\hat{\kappa}_B, \tilde{\kappa}_B ]$, ${\omega}_{B}  \in [\hat{\omega}_{B}, \tilde{\omega}_{B}]$ and ${\lambda}_{E}  \in [\hat{\lambda}_{E}, \tilde{\lambda}_{E}]$.
\end{corollary} 
\begin{proof}
See Appendix \ref{sec:proof_corollary_1}.
\end{proof}

\section{Element subset selection}
 In this section, we introduce ESS that can be used to improve the secrecy performance. {ESS is inspired by transmit antenna selection, which is a technique used to enhance the wireless transmission \cite{5671906,ferdinand2013effects}. } The key idea of ESS is to select the reflecting elements based on IRS-Bob's channel. During data transmission, only the selected elements are turned on, while the other elements are turned off. More specifically, the outdated CSI $\hat{\mathbf{h}}_B$ is used to perform ESS in this work.  The selection is considered random from Eve's point of view, which does not help to improve Eve's SNR but is beneficial to improve Bob's SNR. As a result, the secrecy performance can be effectively enhanced.
 
 As the outdated CSI $\hat{\mathbf{h}}_B$ is known to the BS, the ESS is performed based on its magnitude $| \hat{\mathbf{h}}_B| $. We arrange $| \hat{\mathbf{h}}_B| $ in descending order as 
 \begin{align}
 	\label{eq:RES}
 	| \hat{\mathbf{h}}_{B, s_1}| > | \hat{\mathbf{h}}_{B, s_2}| > ... > | \hat{\mathbf{h}}_{B, s_N}|. 
 \end{align}
 Here, $K$ elements are selected from the total of $N$ reflecting elements, and the set of selected indices is represented by $\mathbf{S} = [s_1,s_2,...,s_K]$. Therefore, the instantaneous SNR at Bob with ESS under outdated and perfect CSI are, respectively, given by  
\begin{align}
	\label{eq:gamma_B_hat_RES}		
	\hat{\gamma}_B^{*}
	& = \frac{PM \beta_{H} \rho^2 K^2}{\sigma_B^2+ PMK\beta_{H}(1-\rho^2)\beta_{B}	 }  \bigg| \frac{1}{K} \sum_{n \in \mathbf{S} }^{} | \hat{\mathbf{h}}_{B,n} |e^{j \tilde {\phi}_n} \bigg|^2, \\
	\label{eq:gamma_B_tilde_RES}		
	\tilde{\gamma}_B^*
	& = \frac{P}{\sigma_B^2} K^2 M \beta_{H}  \bigg|   \frac{1}{K} \big(  \rho \sum_{n \in  \mathbf{S}  }^{} | \hat{\mathbf{h}}_{B,n}| e^{j \Delta {\phi}_n} \! + \! \sum_{n \in  \mathbf{S}  }^{} \mathbf{e}_{B,n} e^{j \tilde {\phi}_n} \big) \bigg| ^2.
\end{align}
Let us define  
 \begin{align} 
	\label{eq:RES_mean}
	\bar{{h}}_B = \frac{1}{K} \sum_{n \in \mathbf{S} }^{} | \hat{\mathbf{h}}_{B,n}| =  \mathbb{E}  \left[  | \hat{\mathbf{h}}_{B, s_1}| , | \hat{\mathbf{h}}_{B, s_2}| , ... , | \hat{\mathbf{h}}_{B, s_K}|  \right].
    \end{align}  
For large number $N$, we obtain $\Pr(x \leq |\hat{\mathbf{h}}_{B,s_K}|) = \frac{N-K}{N}$. It follows that $|\hat{\mathbf{h}}_{B, s_K}| = \sqrt{-\beta_{B} \ln (1-\frac{N-K}{N})} $, since $|\hat{\mathbf{h}}_{B, n}|$ obeys a Rayleigh distribution whose quantile can be computed by  $Q(q)= \sqrt{-\beta_{B} \ln (1-q)}$. Additionally, the pdf and cdf of $ | \hat{\mathbf{h}}_{B, n}| $ are, respectively, given by
\begin{align} f(x) = \frac{2 x}{\beta_{B}}e^{-\frac{x^2}{\beta_B}},  \qquad F(x) = 1- e^{-\frac{x^2}{\beta_B}}.
\end{align}   Thus, the mean value of \eqref{eq:RES_mean} can be calculated  by
\begin{align} 
	\label{eq:RES_mean2}
	\bar{\mu}_B &= \mathbb{E} [\bar{{h}}_B] =  \frac{1}{1-F(|\hat{\mathbf{h}}_{B, s_K}|)} \int_{|\hat{\mathbf{h}}_{B, s_K}|}^{\infty} x f(x) dx \nonumber \\
	= &\frac{1}{1-F(|\hat{\mathbf{h}}_{B, s_K}|)} \int_{|\hat{\mathbf{h}}_{B, s_K}|}^{\infty}  \frac{2 x^2}{\beta_{B}}e^{-\frac{x^2}{\beta_B}} dx \nonumber \\
	= & \frac{ \sqrt{2\beta_{B}}}{1-F(|\hat{\mathbf{h}}_{B, s_K}|)}  \int_{\frac{|\hat{\mathbf{h}}_{B,s_K}|^2}{\beta_{B}}}^{\infty} \sqrt{r} e^{-r} dr  \nonumber \\
	= & \frac{ \sqrt{2\beta_{B}}}{1-F(|\hat{\mathbf{h}}_{B, s_K}|)}  \Gamma\left(  \frac{3}{2}, {\frac{|\hat{\mathbf{h}}_{B,s_K}|^2}{\beta_{B}}}\right),
\end{align} 
{where $\Gamma(\alpha, x) $ is the upper incomplete gamma function defined as $\Gamma(\alpha, x)= \textstyle \int_{x}^{\infty} t^{\alpha-1} e^{-t} d t$, which is available in most software tools.}
According to \cite{mosteller2006some}, when $N$ is large,  $	\bar{{h}}_B$  is asymptotically normally distributed as
\begin{align}
	\label{eq:median_distribution}
	\bar{{h}}_B \sim  \mathcal{AN}\left(   \bar{\mu}_B, \bar{\beta}_B \right)  ,
\end{align}
{where $ \bar{\mu}_B$ is given by \eqref{eq:RES_mean2},} $\bar{\beta}_B =\frac{p(1-p)}{N\left[f\left(F^{-1}(p)\right)\right]^2}$ and $p = 1- e^{\bar{\mu}_{\scriptscriptstyle B}^2/\beta_{\scriptscriptstyle B}}$. Taking advantage of the fact that the random phase error at each element is independent of Bob's channel, we obtain the distribution of $\hat{\gamma}_{B}^*$ and $ \tilde{ \gamma}_{B}^* $ in the following lemma.
\begin{lemma}
	The distributions of Bob's SNR with ESS assuming outdated and perfect CSI at Bob  are, respectively, given by
	\begin{align}
		\label{eq:rec_SNR_dis}	
		\hat{\gamma}_{B}^* \sim \mathrm{Gamma} \left(  \hat{\kappa}_B^*, \hat{\omega}_B^* \right), \\
		\tilde{	\gamma}_{B}^* \sim \mathrm{Gamma} \left(  \tilde{\kappa}_B^*, \tilde{\omega}_B^* \right),
	\end{align}
	where  $
	\hat{\kappa}_B^* = \frac{K\mu_{1}^2\bar{\mu}_B^2 }{2(\bar{\mu}_B^2+\bar{\beta}_B)(1+\mu_{2}-2\mu_{1}^2)+4 K\bar{\beta}_B \mu_{1}^2}$, $
	\tilde{\kappa}_B^* =  \frac{\rho^2 K\mu_{1}^2\bar{\mu}_B^2 }{2\rho^2(\bar{\mu}_B^2+\bar{\beta}_B)(1+\mu_{2}-2\mu_{1}^2)+4 \rho^2K\bar{\beta}_B \mu_{1}^2 + 2(1-\rho^2)\beta_{B}}$, $\hat{\omega}_B^* = \frac{	\mathbb{E}[\hat{\gamma}_B^*]}{\hat{\kappa}_B^*}$ and $\tilde{\omega}_B^* = \frac{	\mathbb{E}[\tilde{\gamma}_B^*]}{\tilde{\kappa}_B^*}$, in which 
	\begin{align}
		\label{eq:average_gamma_B_out}
		\mathbb{E}[\hat{\gamma}_B^*] &=  \frac{PM\beta_{H}\rho^2K (\bar{\mu}_B^2+\bar{\beta}_B) (1-\mu_{1}^2+K\mu_{1}^2)}{	\sigma_B^2+ PMK\beta_{H}(1-\rho^2)\beta_{B} }, \\
		\mathbb{E}[\tilde{\gamma}_B^*] &=  \frac{PM \beta_{H}\rho^2K}{\sigma_{B}^2}\Big((\bar{\mu}_B^2+\bar{\beta}_B) (1-\mu_{1}^2+K\mu_{1}^2)+ \nonumber \\  &\qquad \qquad \qquad \qquad K(1-\rho^2)\beta_{B} \Big),  
	\end{align} 
\end{lemma}

\begin{proof}	
	By using \eqref{eq:median_distribution} and {following the same steps as used in Lemma 1,} we obtain the distribution of $\hat{\gamma}_B^{*}$ and $\tilde{\gamma}_B^{*}$.	{Details are omitted due to space constraints.}	
\end{proof}

\begin{remark} Note that the ESS for Eve is considered random, so the distribution of Eve's SNR with ESS can be easily obtained by Lemma 2, replacing $N$ with $K$. We use $\hat{\gamma}_E^*$ and $\tilde{\gamma}_E^*$ to denote the instantaneous SNR at Eve with ESS assuming outdated and perfect CSI, and we obtain
	\begin{align}
		\label{eq:average_gamma_E_out_RES}
		\hat{\lambda}_{E}^* = \mathbb{E}[\hat{\gamma}_E^*] &= \frac{PM \beta_{H}  \rho^2K {\beta}_{E}}{\sigma_E^2+P MK\beta_H\left(1-\rho^2\right) \beta_E} , \\
		\label{eq:average_gamma_E_perfect_RES}
		\tilde{\lambda}_{E}^* = \mathbb{E}[\tilde{\gamma}_E^*] &= \frac{P}{\sigma_{{E}}^2} M \beta_{H} K {\beta}_{E}.
	\end{align}	
\end{remark}

\begin{remark}
	The SNR distribution at Bob and Eve remain unchanged with ESS. Therefore, the closed-form expression of the SOP with ESS and the corresponding lower bound can be obtained from Theorem 1 and Corollary 1 by setting $\kappa_B \in [\hat{\kappa}_B^*, \tilde{\kappa}_B^* ]$, ${\omega}_{B}  \in [\hat{\omega}_{B}^*, \tilde{\omega}_{B}^*]$ and ${\lambda}_{E}  \in [\hat{\lambda}_{E}^*, \tilde{\lambda}_{E}^*]$.
\end{remark}
Finally, the optimal number of selected reflecting elements can be obtained by 
	\begin{align}
		K_{opt} = \argmin P_{so}^*(K),
			\end{align}
where  $P_{so}^*(K)$ is the SOP as a function of $K$.  Therefore, the SOP with optimal ESS can be calculated by $P_{so}^*(K_{opt})$.

\section{Numerical results}
\vspace{-0.5em}
In this section, we present the numerical results to evaluate the secrecy performance using the proposed ESS method. We obtain the numerical results by averaging over $10^5$ Monte-Carlo simulations. Throughout the simulations, we set $M = 4$, $P = 5$ dBm, $\sigma_{B}^2 = \sigma_{E}^2 = -120$ dBm,  and $N = 100$ (if not
specified otherwise). {We assume that the distance between BS and IRS is $d_1 = 10$ m, and the distance between IRS and Bob/Eve is $d_2 = 80$ m. The pass-loss factors $\beta_{H}$ and $\beta_{B/E}$ are obtained by $\beta_{H} = C_1 d_{1}^{-\alpha_1}$ and $\beta_{B/E} = C_2 d_{2}^{-\alpha_2}$, where $C_1 = -26$ dB, $C_2 = -28$ dB, $\alpha_1=2.2$, and $\alpha_2=3.67$.}
Also, it is assumed that the IRS phase error is uniformly distributed in $[-\frac{\pi}{4},\frac{\pi}{4}]$.

 		\begin{figure*}
	\centering
	\begin{subfigure}[b]{0.45\textwidth}
		\centering
		\includegraphics[width=0.85\linewidth]{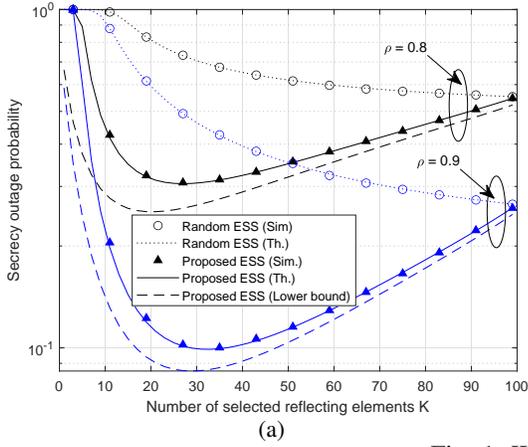} \vspace{-0.5em}
		\caption{}
		\label{fig:iccfig1} \vspace{-0.5em}
	\end{subfigure}
	\hfill
	\begin{subfigure}[b]{0.45\textwidth}
		\centering
		\includegraphics[width=0.85\linewidth]{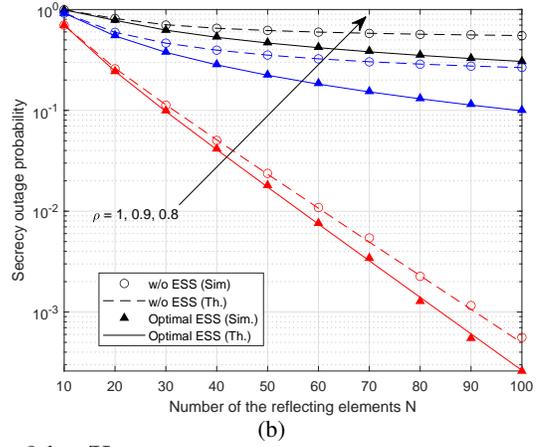}\vspace{-0.5em}
		\caption{}
		\label{fig:iccfig2} \vspace{-0.5em}
	\end{subfigure} \vspace{-0.2em}
	\caption{Worst-case SOP, $R_s = 3$ bps/Hz}\vspace{-1em}
	\label{fig:fig1} 
\end{figure*}
\begin{figure*}
	\centering
	\begin{subfigure}[b]{0.45\textwidth}
		\centering
		\includegraphics[width=0.85\linewidth]{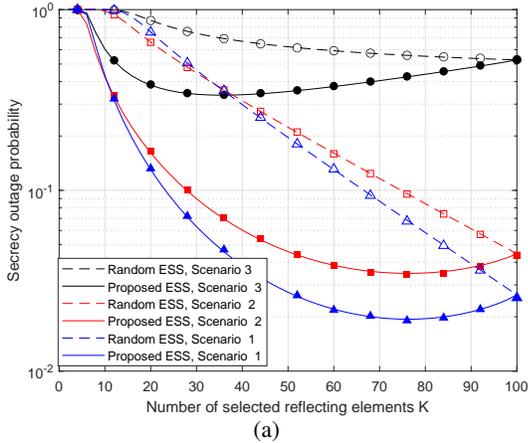}\vspace{-0.5em}
		\caption{}
		\label{fig:iccfig3} \vspace{-0.5em}
	\end{subfigure}
	\hfill
	\begin{subfigure}[b]{0.45\textwidth}
		\centering
		\includegraphics[width=0.85\linewidth]{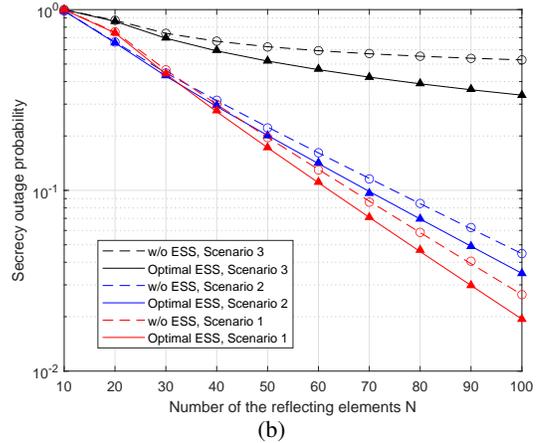}\vspace{-0.5em}
		\caption{}
		\label{fig:iccfig4} \vspace{-0.5em}
	\end{subfigure} \vspace{-0.2em}
	\caption{Comparison of the SOP in different scenarios, $\rho = 0.9$, $R_s = 4$ bps/Hz }\vspace{-1em}
	\label{fig:fig2} 
\end{figure*}
We first evaluate the SOP in the scenario where Bob knows his outdated CSI $\hat{\mathbf{h}}_B$ and Eve knows her perfect CSI ${\mathbf{h}}_E$ in \figref{fig:fig1}. This is the worst case because outdated CSI available at the receiver degrades the SNR at Bob. As a result, the secrecy performance is more critical than in other scenarios. \figref{fig:iccfig1} shows the SOP as a function of the number of selected reflecting elements $K$. Moreover, we compare the performance of the random ESS with that of the proposed ESS. It can be seen that the proposed ESS performs better than the random ESS. Meanwhile, the performance of both selections is the same when $K = 100$, since all elements available are selected. An interesting observation is that the minimum SOP is achieved when only a small number of the reflecting elements are selected.  We also compare the theoretical results with the simulation results. As can be seen, the theoretical results agree very well with the simulation results, which validates Theorem 1. Besides, the lower bound of the proposed ESS behaves similarly to the exact results and therefore can be used to find the optimal number of the selected reflecting elements. \figref{fig:iccfig2} shows the secrecy outage probability as a function of the number of reflecting elements $N$, where we compare the SOP using all elements (w/o ESS) and using the optimal ESS.  It can be seen that the SOP becomes smaller as the number of reflecting elements increases.   Also, the SOP is getting higher as the correlation parameter $\rho$ decreases, which means that the outdated CSI degrades the secrecy performance.  Besides, we find that the secrecy performance with optimal ESS is better than without ESS, especially for large $N$. Moreover, a perfect agreement between the theoretical and simulation results is observed, indicating the correctness of closed-form expressions. 
%

 \vspace{-0.1em}
 In \figref{fig:fig2}, we compare the SOP in different scenarios, where we use curves and markers to represent the theoretical and simulation results, respectively.  The scenarios are defined in section \ref{sec:Secrecy Analysis}.
\figref{fig:iccfig3} illustrates the SOP with the random ESS and the proposed ESS versus the number of selected reflecting elements $K$. It can be seen that the minimum SOP with the proposed ESS for all scenarios is achieved by using only a subset of the IRS. Thus, it is necessary to find the optimal number of selected elements to achieve the minimum SOP. The secrecy performance without ESS and with the optimal ESS is compared in  \figref{fig:iccfig4}, where we can see that the secrecy performance with the optimal ESS  is better than that without ESS, implying that our proposed method can effectively enhance the secrecy performance in all scenarios.

\section{Conclusion}
\label{sec:conclusion} 
In this work, we have studied the secrecy performance in an IRS-assisted multi-antenna BS system. In particular, we assumed that the BS is aware of the outdated CSI and the beamformers at the BS and IRS are designed by the outdated CSI. We characterized the SNR at Bob and Eve, which was used to analyze the SOP. In addition, we proposed an ESS method to improve the secrecy performance.  Both the simulation and analytical results show that the proposed method can effectively improve the secrecy performance. An interesting observation is that the minimum secrecy outage probability can be achieved with a subset of the IRS.


\appendix
\section{Appendices}
\subsection{Proof of Lemma 1}
\label{sec:proof_lemma_1}
 Substituting \eqref{eq:opt_IRS} into \eqref{eq:gamma_B_hat}, 
we obtain $\hat{\gamma}_B = \frac{P}{	\hat{	\sigma}_{B}^2 } M \beta_{H} \rho^2 |  \sum_{n=1}^{N} e^{j \Delta {\phi}_n} | \hat{\mathbf{h}}_{B,n}|   |^2$. Let $\hat{X} = \sum_{n=1}^{N} e^{j \Delta{\phi}_n} | \hat{\mathbf{h}}_{B,n}| $, assuming large number $N$, we have $\re[\hat{X}] \sim \mathcal{N}(\hat{m}_{{u}}, \hat{\delta}_{{u}}^2)$ and $\im[\hat{X}] \sim \mathcal{N}(0, \hat{\delta}_{{v}}^2)$, where $\hat{m}_{{u}} = \frac{N}{2}\sqrt{\pi {\beta_{B}}} \mu_{1}$, $\hat{\delta}_{{u}}^2 =\frac{N}{2}{\beta_{B}} ( 1+\mu_{2}-\frac{\pi}{2} \mu_{1}^2) $, and $\hat{\delta}_{{v}}^2 =\frac{N}{2}{\beta_{B}} ( 1-\mu_{2})$ \cite{badiu2019communication}.  Thus, we obtain $|\hat{X}| ^2 =(\re[\hat{X}])^2 +(\im[\hat{X}])^2   $ follow a Gamma distribution. It follows that $	\hat{\gamma}_{B}$ obeys a Gamma distribution with the mean value $\mathbb{E} [\hat{\gamma}_B] =  \frac{P}{	\hat{	\sigma}_{B}^2 } M \beta_{H} \rho^2 (|\im[\hat{X}]|^2+|\re[\hat{X}]|^2) = \frac{P}{	\hat{	\sigma}_{B}^2 } M \beta_{H} \rho^2 \left( \hat{m}_{{u}}^2+\hat{\delta}_{{u}}^2+\hat{\delta}_{{v}}^2 \right) = \frac{P}{	\hat{	\sigma}_{B}^2 } M \beta_{H} N\beta_{B} \rho^2(1+\frac{\pi}{4}\mu_{1}^2(N-1)) $, and the variance is given by 	$\mathbb{V}[{\hat{\gamma}}_B] = \left( \frac{P}{\hat{	\sigma}_{B}^2 } M \beta_{H} \rho^2 \right) ^2  2\left( 2\hat{m}_{{u}}^2\hat{\delta}_{{u}}^2+\hat{\delta}_{{u}}^4+\hat{\delta}_{{v}}^4\right)$. Therefore, the shape and scale parameter can be computed by
$
\hat{\kappa}_B = \frac{(\mathbb{E}[\hat{\gamma}_B])^2}{\mathbb{V}[{\hat{\gamma}_B}]} = \frac{\left( \hat{m}_{{u}}^2+\hat{\delta}_{{u}}^2+\hat{\delta}_{{v}}^2 \right)^2}{ 2\left( 2\hat{m}_{{u}}^2\hat{\delta}_{{u}}^2+\hat{\delta}_{{u}}^4+\hat{\delta}_{{v}}^4\right)  }  $ and
$
\hat{\omega}_B= \frac{\mathbb{V}[{\hat{\gamma}_B}]}{\mathbb{E}[\hat{\gamma}_B]}
 = \frac{P}{	\hat{	\sigma}_{B}^2 } M \beta_{H} \rho^2 \frac{2\left( 2\hat{m}_{{u}}^2 \hat{\delta}_{{u}}^2+\hat{\delta}_{{u}}^4+\hat{\delta}_{{v}}^4\right) }{\left( \hat{m}_{{u}}^2+\hat{\delta}_{{u}}^2+\hat{\delta}_{{v}}^2 \right)}$, respectively. In the case where the perfect CSI is available, Bob's SNR is given by  $\tilde{\gamma}_B = \frac{P}{	{	\sigma}_{B}^2 } M \beta_{H}  |  \rho \sum_{n=1}^{N} e^{j \Delta {\phi}_n} | \hat{\mathbf{h}}_{B,n}| +  \sum_{n=1}^{N} \mathbf{e}_{B,n} |^2$. Let $\tilde{X} = \sum_{n=1}^{N} e^{j \Delta {\phi}_n} | \hat{\mathbf{h}}_{B,n}| +  \sum_{n=1}^{N} \mathbf{e}_{B,n} $, 
we have $\re[\hat{X}] \sim \mathcal{N}(\tilde{m}_{{u}}, \tilde{\delta}_{{u}}^2)$ 
and $\im[\tilde{X}] \sim \mathcal{N}(0, \tilde{\delta}_{{v}}^2)$, 
where $\tilde{m}_{{u}} = \rho \frac{N}{2}\sqrt{\pi {\beta_{B}}} \mu_{1}$, 
$\tilde{\delta}_{{u}}^2 =\rho^2\frac{N}{2}{\beta_{B}} ( 1+\mu_{2}-\frac{\pi}{2} \mu_{1}^2) +\frac{N}{2}(1-\rho^2)\beta_{B}$, and 
$\tilde{\delta}_{{v}}^2 = \rho^2\frac{N}{2}{\beta_{B}} ( 1-\mu_{2}) + \frac{N}{2}(1-\rho^2)\beta_{B}$. Following the same steps as deriving the scale and shape parameters under the assumption of outdated CSI, we observe $\tilde{\kappa}_B = \frac{\left( \tilde{m}_{{u}}^2+\tilde{\delta}_{{u}}^2+\tilde{\delta}_{{v}}^2 \right)^2}{ 2\left( 2\tilde{m}_{{u}}^2\tilde{\delta}_{{u}}^2+\tilde{\delta}_{{u}}^4+\tilde{\delta}_{{v}}^4\right)  }$ and $\tilde{\omega}_B = \frac{P}{	{	\sigma}_{B}^2 } M \beta_{H}  \frac{2\left( 2\tilde{m}_{{u}}^2 \tilde{\delta}_{{u}}^2+\tilde{\delta}_{{u}}^4+\tilde{\delta}_{{v}}^4\right) }{\left( \tilde{m}_{{u}}^2+\tilde{\delta}_{{u}}^2+\tilde{\delta}_{{v}}^2 \right)} $.

\subsection{Proof of Corollary 1}
\label{sec:proof_corollary_1}
	The SOP in \eqref{eq:SOP} can be rewritten as
$
	P_{so}	=\operatorname{Pr}[\gamma_{\mathrm{B}} \leqslant 2^{R_{\mathrm{s}}}\left(1+\gamma_{\mathrm{E}}\right)-1]  = \operatorname{Pr}[\gamma_{\mathrm{B}} \leqslant 2^{R_{\mathrm{s}}}\gamma_{\mathrm{E}} + \text{constant} ]  
	\geqslant  1- \operatorname{Pr}[\gamma_E / \gamma_B \leqslant 2^{-R_{\mathrm{s}}} ] 
$.
Exploiting the fact that $\gamma_{B}$ and $\gamma_{{E}}$ follow the Gamma and exponential distribution, we have $\frac{ \mathbb{E}[{\gamma}_{B}] \gamma_{E}}{\mathbb{E}[{\gamma}_{E}] \gamma_{B} } $  follows an original Fisher–Snedecor distribution with the degrees of freedom $d_1 =2$, $d_2 = 2 \kappa_B$, {where $\mathbb{E}[{\gamma}_{B}]$ and $\mathbb{E}[{\gamma}_{E}]$ are given by Lemma 1 and Lemma 2, respectively.} As a result, we obtain 
$
	P_{so}  \geqslant 1- \operatorname{Pr}\left[ {\gamma_E}/{\gamma_B} \leqslant 2^{-R_{\mathrm{s}}} \right] =  1- \operatorname{Pr}\left[ \frac{ \mathbb{E}[{\gamma}_{B}] \gamma_{E}}{\mathbb{E}[{\gamma}_{E}] \gamma_{B} } \leqslant 2^{-R_{\mathrm{s}}}\frac{ \mathbb{E}[{\gamma}_{B}] }{\mathbb{E}[{\gamma}_{E}]} \right] 
	=  1- \operatorname{I}_{ \frac{d_1x}{d_1x+d_2}}\left( 1, \kappa_B\right)
	= \left( \frac{ {\lambda}_E}{{2^{-R_{\mathrm{s}}}{\omega}_B} +  {\lambda}_E}\right) ^{\kappa_B}	
$, where $\operatorname{I}$ is the regularized incomplete beta function and $x = 2^{-R_{\mathrm{s}}}\frac{ \mathbb{E}[{\gamma}_{B}] }{\mathbb{E}[{\gamma}_{E}]} $.   
\bibliographystyle{IEEEbib} 
\bibliography{refs}  
\end{document}